\documentclass[%
 reprint,
showpacs,
 amsmath,amssymb,
 aps,
floatfix,
]{revtex4-1}

\usepackage{graphicx}
\usepackage{dcolumn}
\usepackage{bm}

\usepackage{float}
\usepackage{amsfonts}
\usepackage{latexsym}
\usepackage{amssymb}
\usepackage{mathrsfs}
\usepackage{amsmath}
\usepackage{url} 
\usepackage{color}

\begin{document}

\preprint{APS/123-QED}
\title{Multiscale Granger Causality}


\author{L. Faes}
\author{G. Nollo}%
\affiliation{%
 Bruno Kessler Foundation, Trento, Italy
}%
\affiliation{
 BIOtech, Dept. of Industrial Engineering, University of Trento, Italy
}


\author{S. Stramaglia}
\affiliation{
 Dipartimento di Fisica, Universit\'a degli Studi Aldo Moro, Bari, Italy
}%
\affiliation{
 INFN, Sezione di Bari, Italy
}%
\author{D. Marinazzo}
\affiliation{%
 Data Analysis Department, Ghent University, Ghent, Belgium
}%


\date{\today}

\begin{abstract}
In the study of complex physical and biological systems represented by multivariate stochastic processes, an issue of great relevance is the description of the system dynamics spanning multiple temporal scales. While methods to assess the dynamic complexity of individual processes at different time scales are well-established, 
multiscale analysis of directed interactions has never been formalized theoretically, and empirical evaluations are complicated by practical issues such as filtering and downsampling.
Here we extend the very popular measure of Granger causality (GC), a prominent tool for assessing directed lagged interactions between joint processes, to quantify information transfer across multiple time scales.
We show that the multiscale processing of a vector autoregressive (AR) process introduces a moving average (MA) component, and describe how to represent the resulting ARMA process using state space (SS) models and to combine the SS model parameters for computing exact GC values at arbitrarily large time scales. 
We exploit the theoretical formulation to identify peculiar features of multiscale GC in basic AR processes,
and demonstrate with numerical simulations the much larger estimation accuracy of the SS approach compared with pure AR modeling of filtered and downsampled data.
The improved computational reliability is exploited to disclose meaningful multiscale patterns of information transfer between global temperature and carbon dioxide concentration time series, both in paleoclimate and in recent years.

\end{abstract}

\pacs{02.50.Ey, 05.45.Tp, 87.10.Mn, 92.70.Gt }
\maketitle


\section{\label{sec:level1}Introduction}
Granger causality (GC) is a powerful tool for assessing directional interactions from time series data according to the notion of time lagged influence first proposed by Wiener \cite{wiener1956theory} and then formalized by Granger \cite{granger1969} and Geweke \cite{geweke1982, geweke1984} in the framework of vector autoregressive (AR) modeling of stochastic processes. 
Since its formulation, GC has gained increasing popularity and is nowadays ubiquitously employed in several scientific fields ranging from econometrics to social and climate sciences, neuroscience and physiology \cite{freeman1983granger,Smirnov2009,bressler2011,Porta2015}.
The great success of this measure comes from its conceptual simplicity, data-driven nature, and relative ease of implementation. An additional appealing property of GC is the principled interpretation of its generalized probabilistic formulation, which is closely related to the the information-theoretic concept of transfer entropy \cite{barnett2009granger}.

Many processes in physics, biology and other fields exhibit dynamics spanning multiple temporal scales \cite{Ivanov1999461,Thakor2009,Valencia20092202,chou2011wavelet,wang2013multiscale}. The multiscale properties of an observed stochastic process can be explored first resampling at different temporal scales the originally measured realization of the process, and then assessing the dynamical complexity of the rescaled series \cite{costa2002multiscale}.
This approach has been followed with great success to quantify the multiscale behavior of the individual dynamics of scalar processes \cite{Thakor2009,Valencia20092202,chou2011wavelet,wang2013multiscale}. However, its extension to the multiscale computation of the information transfer between processes, though attempted in empirical studies \cite{Lungarella2007,paluvs2014cross}, is far less straightforward. In fact,
the multiscale evaluation of lagged influence measures such as the GC is severely complicated by theoretical and practical issues \cite{barnett2015granger,solo2016state}. These issues arise from the rescaling procedure, which essentially consists in a filtering step eliminating the fast temporal scales (classically performed by averaging \cite{costa2002multiscale}) followed by a downsampling step that coarse-grains the time series around the selected scale. The filtering step leaves theoretically unchanged the GC values, but degrades severely their estimation affecting reliability, stability and data demand \cite{Barnett2011404}. The downsampling step is even more problematic, as it alters GC values in a way that was unknown until very recently \cite{barnett2015granger,solo2016state} and impacts consistently detectability and accuracy of GC estimates.
For these reasons, research on the data-driven inference of the multiscale structure of coupled processes, though holding a great potential, is still largely undeveloped.

In order to provide a formal extension of GC to multiscale analysis, here we introduce for the first time an analytical frame for its computation on linear multivariate stochastic processes subjected to averaging and downsampling. We exploit the theory of state space (SS) models \cite{Aoki1991,barnett2015granger,solo2016state} to yield exact GC values for for coupled processes observed at different time scales. The high computational reliability of the associated multiscale GC estimator is demonstrated in simulated AR processes and then used to explore, spanning a wide range of time scales, the patterns of information transfer between anthropogenic emissions and global temperatures.



\section{Granger Causality}
\subsection{Definition}

To lay the groundwork for multiscale GC computation and introduce notations, we start resuming the calculation of GC for general multivariate processes \cite{granger1969,geweke1984}. 
Let us consider a discrete-time, stationary vector stochastic process  composed of \textit{M} real-valued zero-mean scalar processes, $Y_n=[y_{1,n}\cdots y_{M,n}]^T$, $-\infty < n < \infty$, and assume the process $y_{j}$ as the \textit{target} and the process $y_{i}$ as the \textit{driver}
(the remaining $M-2$ processes form the vector $Y_k$, where $k=\{1,\ldots,M \} \backslash \{i,j \}$).
Then, denoting the present and the past of vector and scalar variables respectively as $Y_n$, $y_n$, and $Y_n^- = [Y_{n-1}^T Y_{n-2}^T \cdots]$, $y_n^-=[y_{n-1}y_{n-2}\cdots]$, GC from $y_{i}$ to $y_{j}$ (conditional on $Y_k$) quantifies the extent to which $y_{i,n}^-$ improves the prediction of $y_{j,n}$ above and beyond the extent to which $y_{j,n}$ is predicted by $y_{j,n}^-$ and $Y_{k,n}^-$.
This definition is assessed in the time domain performing a regression of the present of the target on the past of all processes, yielding the prediction error $e_{j|ijk,n}=y_{j,n}-\mathbb{E}[y_{j,n}|Y_n^-]$, and on the past of all processes except the driver, yielding the prediction error $e_{j|jk,n}=y_{j,n}-\mathbb{E}[y_{j,n}|y_{j,n}^-,Y_{k,n}^-]$ ($\mathbb{E}$ is the expectation operator).
The prediction error variances resulting from these "full" and "restricted" regressions, $\lambda_{j|ijk}=\mathbb{E}[e_{j|ijk,n}^2]$ and $\lambda_{j|jk}=\mathbb{E}[e_{j|jk,n}^2]$ are then combined to yield GC from $y_{i}$ to $y_{j}$ as \cite{geweke1984}
\begin{equation} \label{eq:GC}
		F_{i\rightarrow j} = \ln \frac {\lambda_{j|jk}}{\lambda_{j|ijk}} .
\end{equation}
The measure (\ref{eq:GC}) is the log-likelihood ratio for the two linear regressions associated with the projections $\mathbb{E}[y_{j,n}|Y_n^-]$ and $\mathbb{E}[y_{j,n}|y_{j,n}^-,Y_{k,n}^-]$ \cite{barnett2015granger}. According to the axiomatic definition of transfer entropy and its equivalence (up to a factor 2) to Eq.(\ref{eq:GC}) for Gaussian variables \cite{barnett2009granger}, the GC can be interpreted as the rate of "information transfer" from driver to target.

\subsection{State-Space Formulation}
Following the derivations of a recent work by Barnett et al. \cite{barnett2015granger}, now we move to describe the computation of GC for state state space (SS) processes.
The very well known linear SS representation of an observed multivariate process $Y$ is given by \cite{anderson1979optimal}
\begin{subequations} \label{eq:eqSS}
\begin{align}
		X_{n+1} &= \mathbf{A} X_{n} + W_{n} \label{eq:eqSSstate} \\
		Y_n &= \mathbf{C} X_{n} + V_{n} \label{eq:eqSSobs}
\end{align}
\end{subequations}
where $X$ is the state (unobserved) process, and $W$ and $V$ are zero-mean white noise processes with covariances {\boldmath$\Xi$}$\equiv$$\mathbb{E}[W_nW_n^T]$ and {\boldmath$\Psi$}$\equiv$$\mathbb{E}[V_nV_n^T]$, and cross-covariance {\boldmath$\Theta$}$\equiv$$\mathbb{E}[W_nV_n^T]$.

The SS process has an equivalent representation, referred to as "innovations form" SS (ISS), evidencing the \textit{innovations} $E_n=Y_n-\mathbb{E}[Y_n|Y_n^-]$, i.e., the residuals of the linear regression of $Y_n$ on its infinite past $Y_n^-$, whose covariance matrix is {\boldmath$\Phi$}$\equiv$$\mathbb{E}[E_nE_n^T]$. The ISS representation, which is typically associated with Kalman filtering, is characterized by the state process $Z_n=\mathbb{E}[X_n|Y_n^-]$ and by the Kalman Gain matrix $\mathbf{K}$:
\begin{subequations} \label{eq:eqISS}
\begin{align}
		Z_{n+1} &= \mathbf{A} Z_{n} + \mathbf{K} E_{n} \label{eq:eqISSstate} \\
		Y_n &= \mathbf{C} Z_{n} + E_{n} . \label{eq:eqISSobs}
\end{align}
\end{subequations}
The SS and ISS representations share the state and observation matrices $\mathbf{A}$ and $\mathbf{C}$, and differ in the noise matrices  ($\mathbf{\Xi},\mathbf{\Psi},\mathbf{\Theta}$) and ($\mathbf{K}, \mathbf{\Phi}$).
To find the ISS parameters ($\mathbf{A},\mathbf{C},\mathbf{K},\mathbf{\Phi}$) from the SS parameters 
($\mathbf{A},\mathbf{C},\mathbf{\Xi},\mathbf{\Psi},\mathbf{\Theta}$) it is necessary to solve a so-called discrete algebraic Ricatti equation (\textit{DARE}), formulated in terms of the state error variance matrix $\mathbf{P}$:
\begin{equation} \label{eq:DARE}
\begin{aligned}
	\mathbf{P} &= \mathbf{A}\mathbf{P}\mathbf{A}^T + \mathbf{\Xi} \\
						&-(\mathbf{A}\mathbf{P}\mathbf{C}^T+\mathbf{\Theta})
				(\mathbf{C}\mathbf{P}\mathbf{C}^T+\mathbf{\Psi})^{-1} (\mathbf{C}\mathbf{P}\mathbf{A}^T+\mathbf{\Theta}^T),
\end{aligned}
\end{equation}
from which $\mathbf{K}$ and $\mathbf{\Phi}$ are obtained as
\begin{equation} \label{eq:SStoISS}
\begin{aligned}
						\mathbf{\Phi} &= \mathbf{C}\mathbf{P}\mathbf{C}^T + \mathbf{\Psi} \\
						\mathbf{K} &= (\mathbf{A}\mathbf{P}\mathbf{C}^T + \mathbf{\Theta})\mathbf{\Phi}^{-1}.
\end{aligned}
\end{equation}

Then, GC can be computed from the ISS parameters as follows \cite{barnett2015granger}.
The error variance of the full regression is simply the $j-th$ diagonal element of the innovation covariance,  $\lambda_{j|ijk}=\mathbf{\Phi}(j,j)$. 
The error of the restricted regression is obtained by forming a $submodel$ that excludes the driver process, i.e. considering a state space model with state equation (\ref{eq:eqISSstate}) and observation equation 
\begin{equation} \label{eq:SSreduced}
		Y_n^{(jk)}=\mathbf{C}^{(jk)} Z_n + E_n^{(jk)}
\end{equation}
where the superscript $^{(a)}$ denotes selection of the rows with indices $a$ of a matrix.
Of note, this technique represents one of the key elements in the derivation of a rigorous formalism for defining the information flow between the states of both discrete stochastic mappings and continuous time stochastic systems \cite{XSLiang2016}. Here, we have that
the submodel (\ref{eq:eqISSstate}, \ref{eq:SSreduced}) is an SS model with parameters ($\mathbf{A},\mathbf{C}^{(jk)},\mathbf{K}\mathbf{\Phi}\mathbf{K}^T, \mathbf{\Phi}(jk,jk),\mathbf{K}\mathbf{\Phi}(:,jk)$), which can be converted to an ISS model with innovation covariance $\mathbf{\Phi}^R$ solving the \textit{DARE} (\ref{eq:DARE},\ref{eq:SStoISS}), so that the restricted error variance becomes $\lambda_{j|jk}=\mathbf{\Phi}^R(j,j)$. This shows that GC can be computed numerically from the ISS parameters ($\mathbf{A},\mathbf{C},\mathbf{K},\mathbf{\Phi}$) of an observed process $Y$.

\section{Multiscale Granger Causality}
In this Section we develop our framework for the multiscale computation of GC for linear multivariate processes. Here we consider the most common operalization of GC, i.e. that grounded on the AR representation of multivariate processes \cite{granger1969,barnett2009granger,bressler2011,Porta2015}:
\begin{equation} \label{eq:VAR}
Y_n = \sum_{k=1}^{p}{\mathbf{A}_k Y_{n-k} + U_n},
\end{equation}
where $p$ is the model order, $A_k$ are $M\times M$  matrices of coefficients, and
$U_n=[u_{1,n}\cdots u_{M,n}]^T$ is a vector of $M$ zero mean Gaussian innovation processes with
covariance matrix {\boldmath$\Sigma$}$\equiv$$\mathbb{E}[U_nU_n^T]$.
To study the observed process $Y$ at the temporal scale identified by the scale factor $\tau$, we apply to each constituent process $y_m, m=1,\ldots,M$,the following transformation which performs a weighted average of $q$ consecutive samples of the process:
\begin{equation} \label{eq:MSY}
\bar{y}_{m,n} = \sum_{l=0}^{q}{b_{l} y_{m,n\tau-l}}.
\end{equation}
This rescaling operation corresponds to transform the original process $Y$ through a two-step procedure that consists of the following \textit{filtering} and \textit{downsampling} steps, 
performed respectively with a filter of order $q$ and a rate of downsampling equal to $\tau$. The filtering and downsampling steps yield
respectively the processes $\tilde{Y}$ and $\bar{Y}$ defined as:
\begin{subequations} \label{eq:AvgDws}
\begin{align}
		\tilde{Y}_n &= \sum_{l=0}^{q}{b_l Y_{n-l}} , \label{eq:Avg} \\
		\bar{Y}_n &= \tilde{Y}_{n\tau} , n=1,\ldots,N/\tau .\label{eq:Dws}
\end{align}
\end{subequations}
The change of scale in (\ref{eq:MSY}) generalizes the averaging procedure originally proposed in \cite{costa2002multiscale}, which sets $q=\tau-1$ and $b_l=1/\tau$. In this study we identify the $b_l$ as the coefficients of a linear lowpass filter with cutoff frequency set at $f_{\tau}=1/2\tau$ to avoid aliasing in the subsequent downsampling step.
Here, we develop a FIR filter of order $q$ using the window method and implementing a Hamming window \cite{oppenheimdigital}. The design of a causal filter which performs one-side filtering was chosen on purpose to avoid that past and future samples mix up in the filtering process with potentially harmful consequences on the evaluation of causality. The use of one-side filtering is in agreement with the adoption of a Euler forward scheme, rather than a central differencing scheme, in the definition of causal measures of information flow (see, e.g., \cite{XSLiang2014,XSLiang2015}). Moreover, the use of a FIR filter achieves better elimination of the fast temporal scales with respect to the averaging procedure commonly adopted in multiscale complexity analysis \cite{costa2002multiscale}; in our preliminary work \cite{faes2016multiscale} we have shown indeed that the use of simple averaging may induce the detection of spurious causal influences over uncoupled directions.


Substituting (\ref{eq:VAR}) in (\ref{eq:Avg}), the filtering step leads to the process representation:
\begin{equation} \label{eq:VARMAAvg}
\tilde{Y}_n = \sum_{k=1}^{p}{\mathbf{A}_k \tilde{Y}_{n-k}} + \sum_{l=0}^{q}{\mathbf{B}_l U_{n-l}}
\end{equation}
where $\mathbf{B}_l= b_l \mathbf{I}_M$ ($\mathbf{I}_M$ is the $M\times M$ identity matrix). Hence, the change of scale introduces a moving average (MA) component of order $q$ in the original AR$(p)$ process, transforming it into an ARMA$(p,q)$ process. 
Then, exploiting the close relation between ARMA and SS models \cite{Aoki1991}, the process (\ref{eq:VARMAAvg}) is turned into an ISS model by  defining the state process $\tilde{Z}_n=[Y_{n-1}^T \cdots Y_{n-p}^T U_{n-1}^T \cdots U_{n-q}^T]^T$ that, together with $\tilde{Y}_n$, obeys the state equations
(\ref{eq:eqISS}) with parameters ($\tilde{\mathbf{A}},\tilde{\mathbf{C}},
\tilde{\mathbf{K}},\tilde{\mathbf{\Phi}}$), where
\[\tilde{\mathbf{C}}
=
\begin{bmatrix}
	\mathbf{A}_1&\cdots&\mathbf{A}_p & \mathbf{B}_1&\cdots&\mathbf{B}_{q}
\end{bmatrix},
\]
\[\tilde{\mathbf{A}}
=
\begin{bmatrix}
          \tilde{\mathbf{C}}& & \\
		\mathbf{I}_{M(p-1)}&\mathbf{0}_{M(p-1)\times M(q+1)}& \\
		\mathbf{0}_{M\times M(p+q)}&      & \\
		\mathbf{0}_{M(q-1)\times Mp}&\mathbf{I}_{M(q-1)}&\mathbf{0}_{M(q-1)\times M}
\end{bmatrix},
\]
\[\tilde{\mathbf{K}}
=
\begin{bmatrix}
	\mathbf{I}_M & \mathbf{0}_{M\times M(p-1)} &\mathbf{B}_0^{-T} & \mathbf{0}_{M\times M(q-1)}
\end{bmatrix}^T,
\]
and where $\tilde{\mathbf{\Phi}} = \mathbf{B}_0 \mathbf{\Sigma} \mathbf{B}_0^T$ is the covariance of the innovations $\tilde{E}_n=\mathbf{B}_0 U_n$.
Moreover, the downsampled process $\bar{Y}_n$ can be put in ISS form directly from the ISS formulation of the filtered process $\tilde{Y}_n$: exploiting a recent result (theorem III in \cite{solo2016state}), we find that $\bar{Y}_n=\tilde{Y}_{n\tau}$ has an ISS representation with state process
$\bar{Z}_n=\tilde{Z}_{n\tau}$, innovation process $\bar{E}_n=\tilde{E}_{n\tau}$, and parameters
($\bar{\mathbf{A}},\bar{\mathbf{C}},\bar{\mathbf{K}},\bar{\mathbf{\Phi}}$), where 
$\bar{\mathbf{A}}=\tilde{\mathbf{A}}^\tau$, $\bar{\mathbf{C}}=\tilde{\mathbf{C}}$, and where 
$\bar{\mathbf{K}}$ and $\bar{\mathbf{\Phi}}$ are obtained solving the \textit{DARE} (\ref{eq:DARE},\ref{eq:SStoISS}) for the SS model ($\bar{\mathbf{A}},\bar{\mathbf{C}},\mathbf{\Xi}_\tau,
\tilde{\mathbf{\Phi}},\mathbf{\Theta}_\tau$) with
\begin{equation} \label{eq:QSdown}
\begin{aligned}
	\mathbf{\Theta}_\tau &= \tilde{\mathbf{A}}^{\tau-1} \tilde{\mathbf{K}}\tilde{\mathbf{\Phi}} \\
	\mathbf{\Xi}_\tau &= \tilde{\mathbf{A}} \mathbf{\Xi}_{\tau-1} \tilde{\mathbf{A}}^T
	+ \tilde{\mathbf{K}}\tilde{\mathbf{\Phi}}\tilde{\mathbf{K}}^T, \tau\geq 2 \\
	\mathbf{\Xi}_1 &= \tilde{\mathbf{K}} \tilde{\mathbf{\Phi}} \tilde{\mathbf{K}}^T, \tau=1 .
\end{aligned}
\end{equation}
\begin{figure}[ht]
\includegraphics{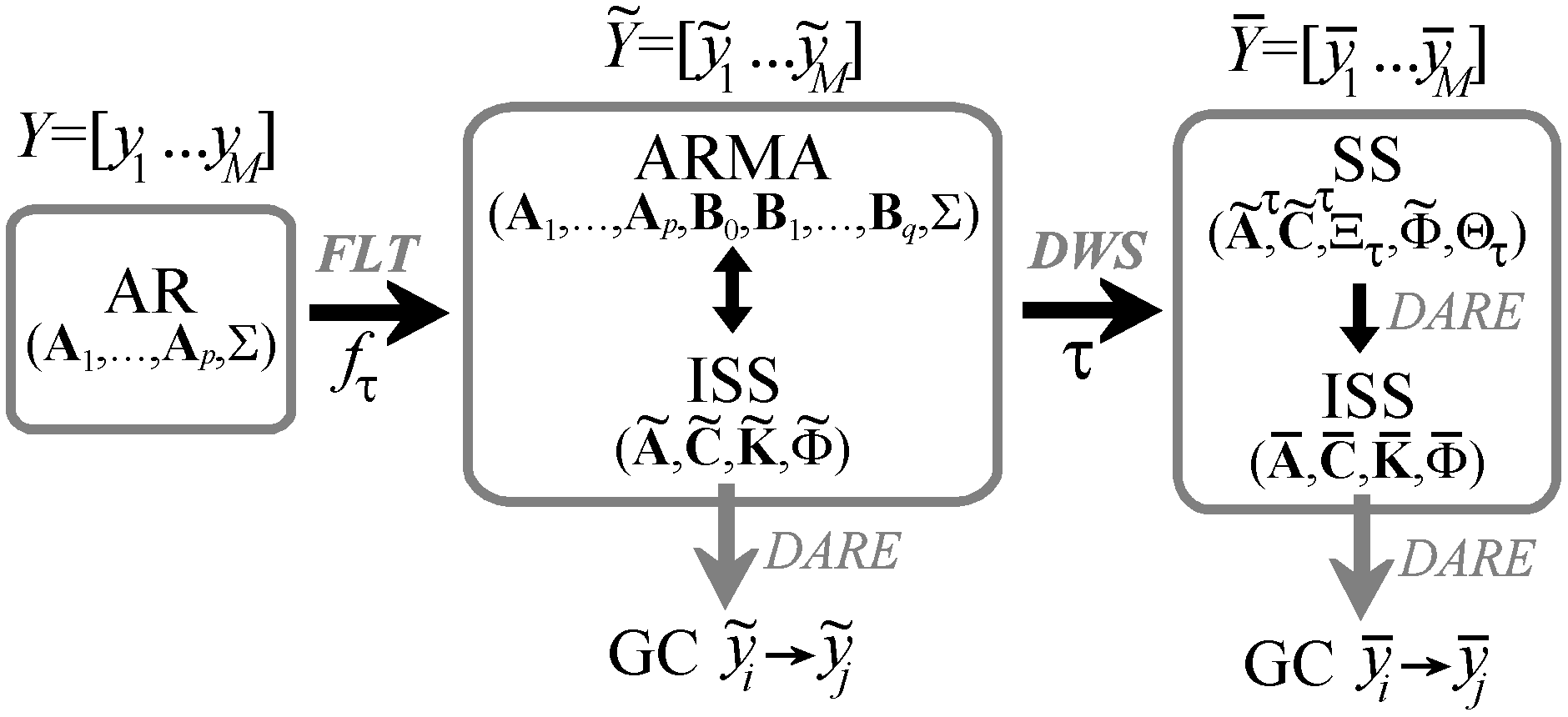}
\caption{\label{fig:scheme} Schematic representation of a linear multivariate AR process (left) and of its multiscale representation obtained through filtering (FLT) and downsampling (DWS) steps. 
At each step, an innovation state space (ISS) model can be defined which describes the multivariate process; then ISS submodels can be formed and, after solving a discrete algebraic Ricatti Equation (DARE), GC can be computed for any scale factor $\tau \geq 1$ ($\tau=1$ yields GC for the non-rescaled processes).}
\end{figure}

The overall procedure for multiscale analysis is depicted in Fig.~\ref{fig:scheme}: filtering with cutoff $f_{\tau}$ the AR($p$) process $Y$ yields an ARMA($p,q$) process, which is equivalent to an ISS process; the subsequent downsampling yields a different SS process, which in turn can be converted to the ISS form solving the \textit{DARE}.
Thus, both filtered and downsampled processes are described by ISS models, whose parameters can be used to compute GC by forming a submodel in which the target is observed without considering the driver process (eq. (\ref{eq:SSreduced})) and solving the \textit{DARE} for this submodel. This procedure allows analytical computation of GC measures for multiscale (filtered and downsampled) processes, which is illustrated in the following for simulated and real time series.

\section{Simulation Study} \label{sec:simu}
\subsection{VAR process with time delayed causal interactions}
\begin{figure}[ht]
\includegraphics[width=8.8 cm]{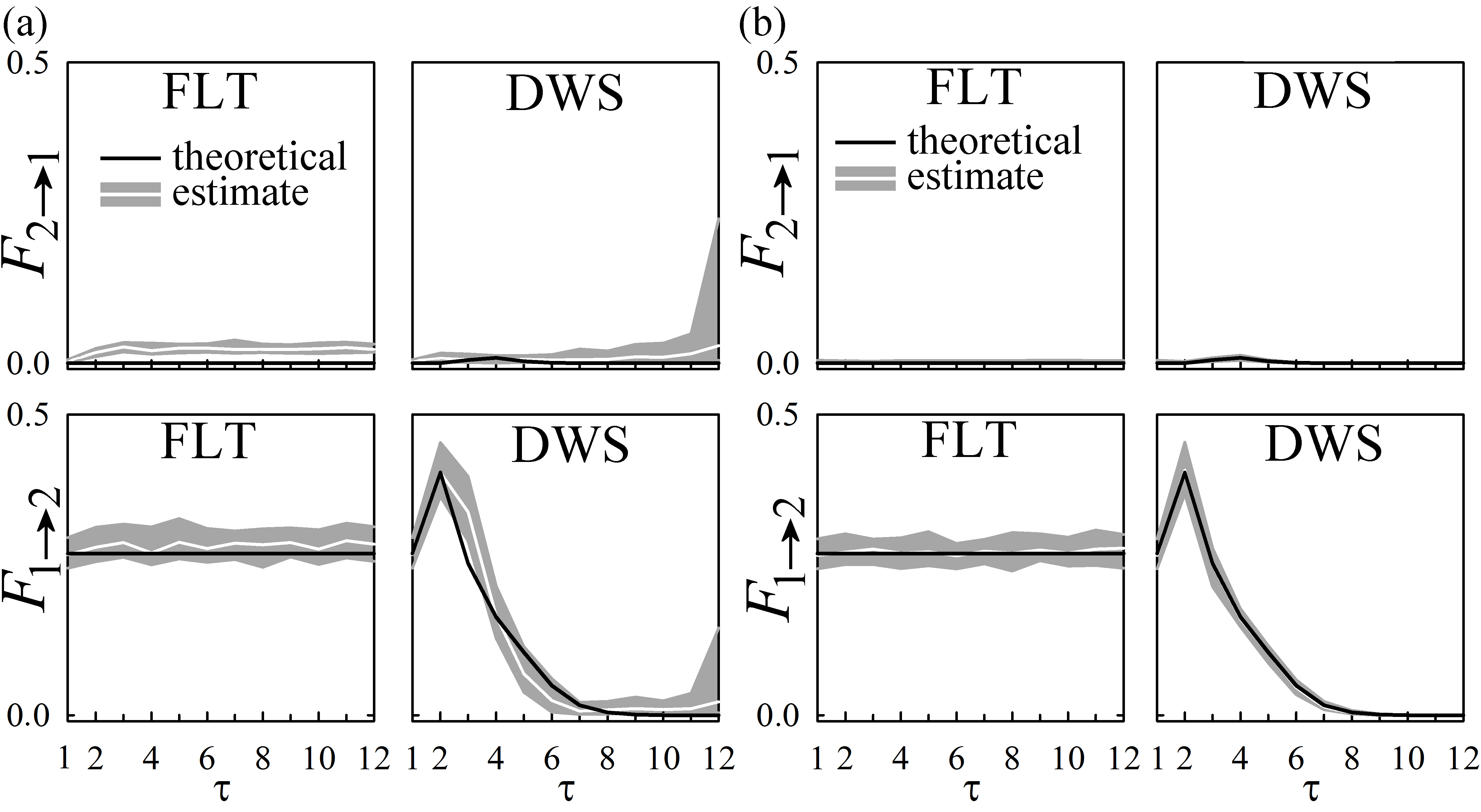}
\caption{\label{fig:simu1} Multiscale GC analysis of the AR process (\ref{eq:simuVAR}) configured to unidirectional coupling from $y_1$ to $y_2$. Plots depict the theoretical values (black lines) and distribution of estimates (median: white lines; interquartile range: grey areas) of GC computed as a function of the time scale after filtering (FLT) and downsampling (DWS). Estimates are obtained using the na\"{\i}ve AR approach (a) and the proposed framework implemented using a lowpass FIR filter of order $q=6$ (b).}
\end{figure}

\begin{figure}[ht]
\includegraphics[width=8.8 cm]{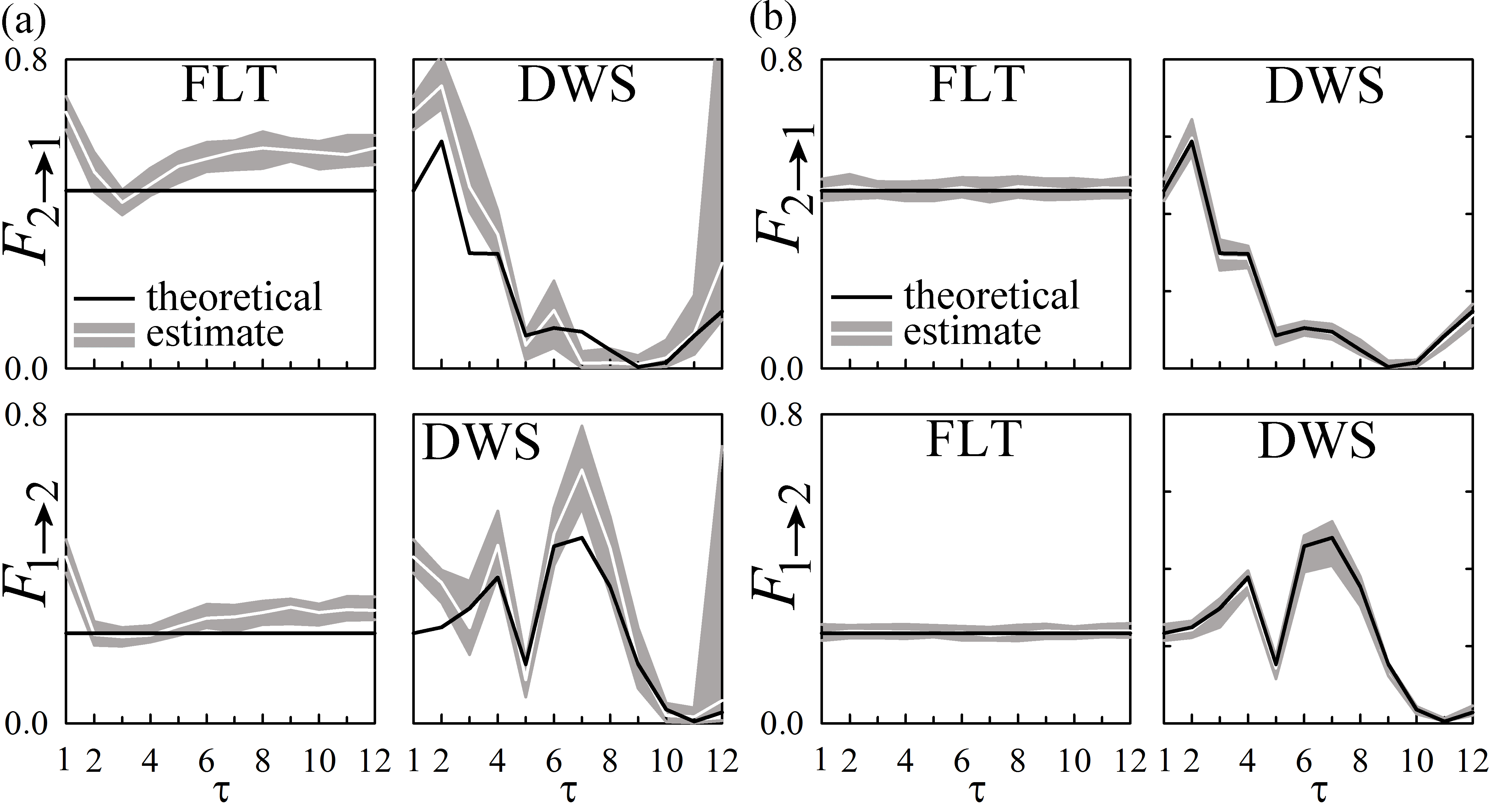}
\caption{\label{fig:simu2} Multiscale GC analysis of the AR process (\ref{eq:simuVAR}) configured to bidirectional coupling between $y_1$ and $y_2$. Plots and symbols are as in Fig.~\ref{fig:simu1}.}
\end{figure}

Theoretical analysis and simulations are first performed for the bivariate AR process with equations:
\begin{subequations} \label{eq:simuVAR}
\begin{align}
		y_{1,n} &= c_{11} y_{1,n-d_{11}} + c_{12} y_{2,n-d_{12}} + u_{1,n} \\
		y_{2,n} &= c_{22} y_{2,n-d_{22}} + c_{21} y_{1,n-d_{21}} + u_{2,n}
\end{align}
\end{subequations}
with iid noise processes $u_{1,n}, u_{2,n} \sim \mathcal{N}(0,1)$. The parameters in (\ref{eq:simuVAR}) are set to generate autonomous dynamics with strength $c_{ii}$ and lag $d_{ii}$ for each scalar process $y_i$, and causal interactions with strength $c_{ij}$ and lag $d_{ij}$ from $y_j$ to $y_i$
($i,j=1,2$).
We consider two parameter configurations: unidirectional interaction at lag 2 from $y_1$ to $y_2$, obtained setting $c_{12}=0$ and $c_{21}=0.5, d_{21}=2$, where also autonomous dynamics are generated for $y_1$ but not for $y_2$ ($c_{11}=0.5, d_{11}=1, c_{22}=0$); bidirectional interactions with different lags and strengths ($c_{12}=0.75, d_{12}=2; c_{21}=0.5, d_{21}=7$) in the presence of autonomous dynamics for both processes ($c_{11}=c_{22}=0.5, d_{11}=d_{22}=1$).
First, we study the exact values of multiscale GC obtained from the true AR parameters. The theoretical trends depicted in Figs.~\ref{fig:simu1} and ~\ref{fig:simu2} (black solid lines) document from the perspective of SS modeling the  invariance of GC under filtering, already proven in \cite{Barnett2011404}. The behavior of the information transfer across multiple temporal scales is thus shaped by the downsampling step, revealing the tendency of GC to peak at scales corresponding with the lag of the imposed causal interactions: maximal information transfer is found at $\tau=2$ for $F_{1\rightarrow2}$ in the unidirectional scheme (Fig.~\ref{fig:simu2}a,b), and at $\tau=7$ for $F_{1\rightarrow2}$ and $\tau=2$ for $F_{2\rightarrow1}$ in the bidirectional scheme (Fig.~\ref{fig:simu2}c). The behavior is general, in the sense that it was observed also for different parameter configurations.

Next, we test reliability of multiscale GC estimates obtained from finite length realizations of (\ref{eq:simuVAR}) (a) following a na\"{\i}ve approach whereby GC is computed performing full and restricted regressions on the filtered and downsampled time series, and (b) performing AR identification on the original time series and then applying the new proposed framework to the estimated AR parameters \cite{[in this study AR models were identified through ordinary least squares and using the Bayesian Information Criterion (BIC) to set the model order as seen e.g. in ]marple1987digital}.
Application of the two approaches to 100 process realizations of 500 points is depicted in Figs.~\ref{fig:simu1}a,~\ref{fig:simu2}a and in Figs.~\ref{fig:simu1}b,~\ref{fig:simu2}b respectively, and indicates the need of state-space analysis: while the computation of GC after filtering and downsampling returns strongly biased and highly variable estimates, the new framework yields accurate detection of GC across multiple scales.

\subsection{VAR process with multiscale causal interactions}
As a second simulation example, we consider the case of coupled stochastic processes displaying multiscale structure and scale-dependent causal interactions. Specifically, we consider the bivariate process $Y_n=[y_{1,n} y_{2,n}]^T$ obtained as the instantaneous mixing of pairs of scalar processes taken from the two bivariate AR processes $X_n=[x_{1,n} x_{2,n}]^T$ and $Z_n=[z_{1,n} z_{2,n}]^T$ :
\begin{subequations} \label{eq:simuVARmix}
\begin{align}
		x_{1,n} &= 1.9 x_{1,n-1} - 0.9025 x_{1,n-2} + u_{1,n} \\
		x_{2,n} &= 0.5 x_{1,n-1} + u_{2,n} , \\
		z_{1,n} &= 1.6929 z_{1,n-1} - 0.9025 z_{1,n-2} + w_{1,n} \\
		z_{2,n} &= z_{1,n-1} + w_{2,n} , \\
		y_{1,n} &= x_{1,n} + z_{2,n} \\
		y_{2,n} &= x_{2,n} + z_{1,n} .
\end{align}
\end{subequations}
In the bivariate processes $X$ and $Z$, autonomous dynamics are set for the subprocesses $x_1$ and $z_1$ according to Eqs. (\ref{eq:simuVARmix}a, \ref{eq:simuVARmix}c); the autonomous rhythms are obtained by placing a pole with modulus $\rho_{x_1}=\rho_{z_1}=0.95$ in the complex plane representation of each individual subprocess, and the phase of the poles is varied to obtain slow oscillations for $x_1$ ($\phi_{x_1}=0$) and faster oscillations for $z_1$ ($\phi_{z_1}=0.47$).
Moreover, causal interactions are imposed from $x_1$ to $x_2$ and from $z_1$ to $z_2$ according to Eqs. (\ref{eq:simuVARmix}b, \ref{eq:simuVARmix}d); in this study, the variances of the uncorrelated innovations are set to $\lambda_{u_1}=0.25, \lambda_{u_2}=0.5,$ for the process $U$, and to $\lambda_{w_1}=1, \lambda_{w_2}=0.5,$ for the process $W$. Then, the mixing obtained with Eqs. (\ref{eq:simuVARmix}e, \ref{eq:simuVARmix}f) is such that the bivariate process $Y$ exhibits causal interactions from $y_2$ to $y_1$ visible at small time scales for the faster oscillations, as well as causal interactions  from $y_1$ to $y_2$ visible at larger time scales for the slower oscillations.

\begin{figure}[ht]
\centering
\includegraphics[width=8.7 cm]{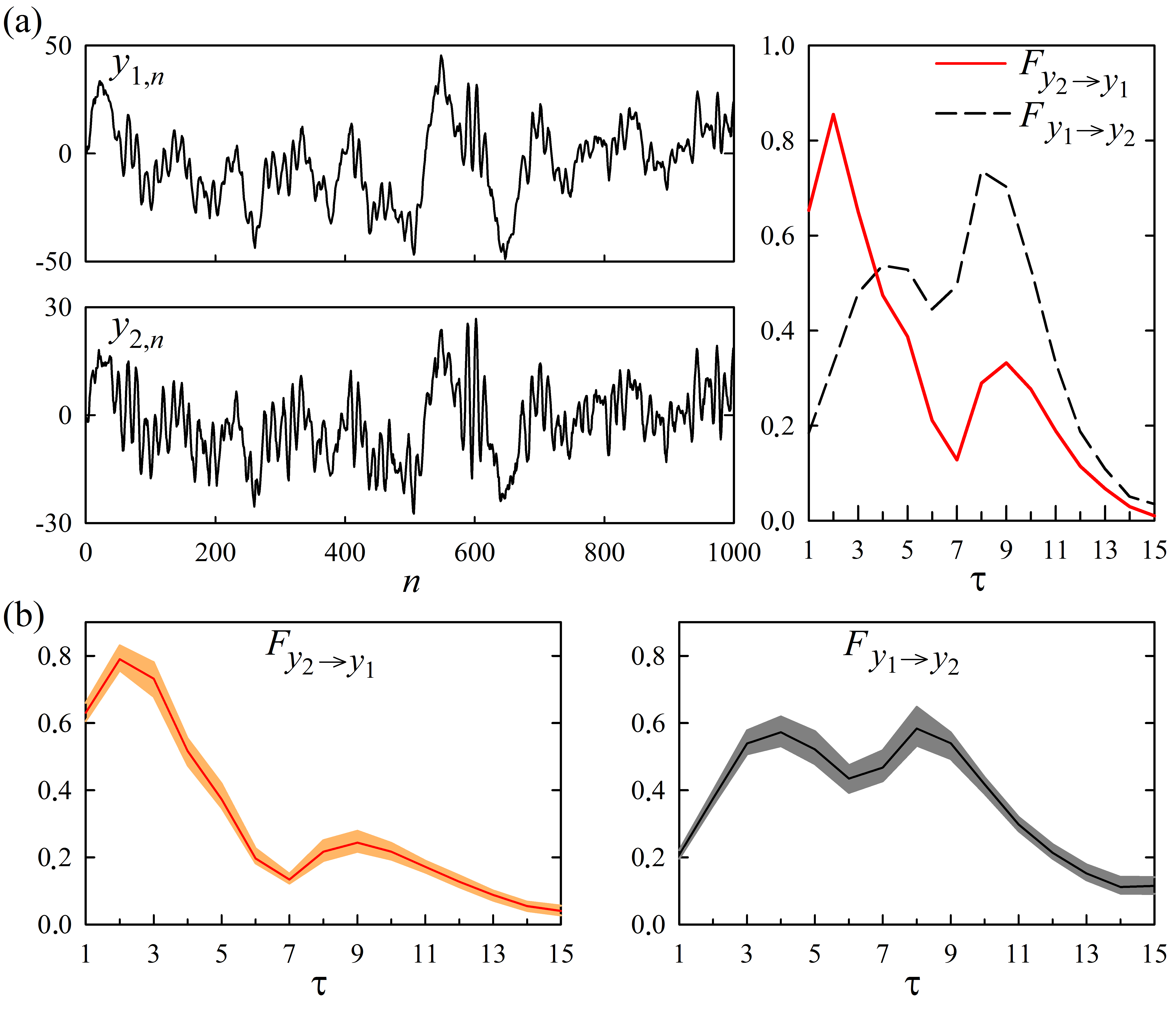}
\caption{\label{fig:Simu_Rev} Multiscale GC analysis of the bivariate process $Y$ defined as in Eq. (\ref{eq:simuVARmix}). Plots depict two exemplary realizations of the observed process $Y_n=[y_{1,n} y_{2,n}]$ together with the GC computed as a function of the time scale $\tau$ along the two directions of interaction (a) and the distribution of estimates (median: solid lines; interquartile range: shaded areas) obtained over 100 realizations of the process (b).}
\end{figure}

To test the ability of our framework to detect these multiscale behaviors, we performed VAR identification on realizations of 1000 data points of the observed process $Y$ and then computed the GC for temporal scales ranging from 1 to 15; the model order was optimized using the BIC criterion, and the order of the lowpass FIR filter was set to $q=6$. The results of the analysis are reported in Fig. \ref{fig:Simu_Rev}. The exemplary realizations shown in Fig. \ref{fig:Simu_Rev}(a) display multiscale patterns characterized by a slow rhythm (cycle of $\sim 70$ points) superimposed to faster oscillations (cycle of $\sim 13$ points);
the multiscale GC analysis reveals that the direction of interaction is from $y_2$ to $y_1$ for the fast oscillations ($F_{y_2\to y_1}$ is maximum at low time scales),
and from $y_1$ to $y_2$ for the slower rhythm ($F_{y_1 \to y_2}$ emerges at higher time scales when fast oscillations are filtered out). These results are confirmed by the analysis extended to several process realizations reported in Fig. \ref{fig:Simu_Rev}(b), which indicates that GC peaks at scale $\tau=2$ along the direction $y_2 \to y_1$, and at scales $\tau=4$ and $\tau=8$ along the direction $y_1 \to y_2$, thus detecting the multiscale patterns of bidirectional interaction imposed in the simulation.

\section{Practical Application} \label{sec:appl}
As a practical application, we consider the multiscale analysis of GC between carbon dioxide concentration ($CO_2$) and global temperature ($GT$). It is widely considered that the raise of $CO_2$ is a main  cause  of  global  warming \cite{Booth2012}, although  the validity of such causal relation is still under debate. The problem of understanding  the causes of climate change is usually tackled by numerical experiments using Global Climate Models \cite{Delworth2006} which aim at catching the complexity of climate dynamics. However, data-driven approaches, as GC, are also fruitful in assessing cause-effect relationships between temperature and external forcings. In \cite{Kodra2011} it has been shown that $CO_2$ Granger causes temperature, based on data from 1860 to 2008, partly from ice cores, and analyzing second differences of both $CO_2$ and $GT$. Similar conclusions were found in \cite{Attanasio2012}, using GC,  in \cite{Stips2016} by estimating the time rate of information flowing from one time series to the other, and in \cite{Smirnov2009} using a physical approach.

\begin{figure}[ht!]
\includegraphics [width=8.5 cm] {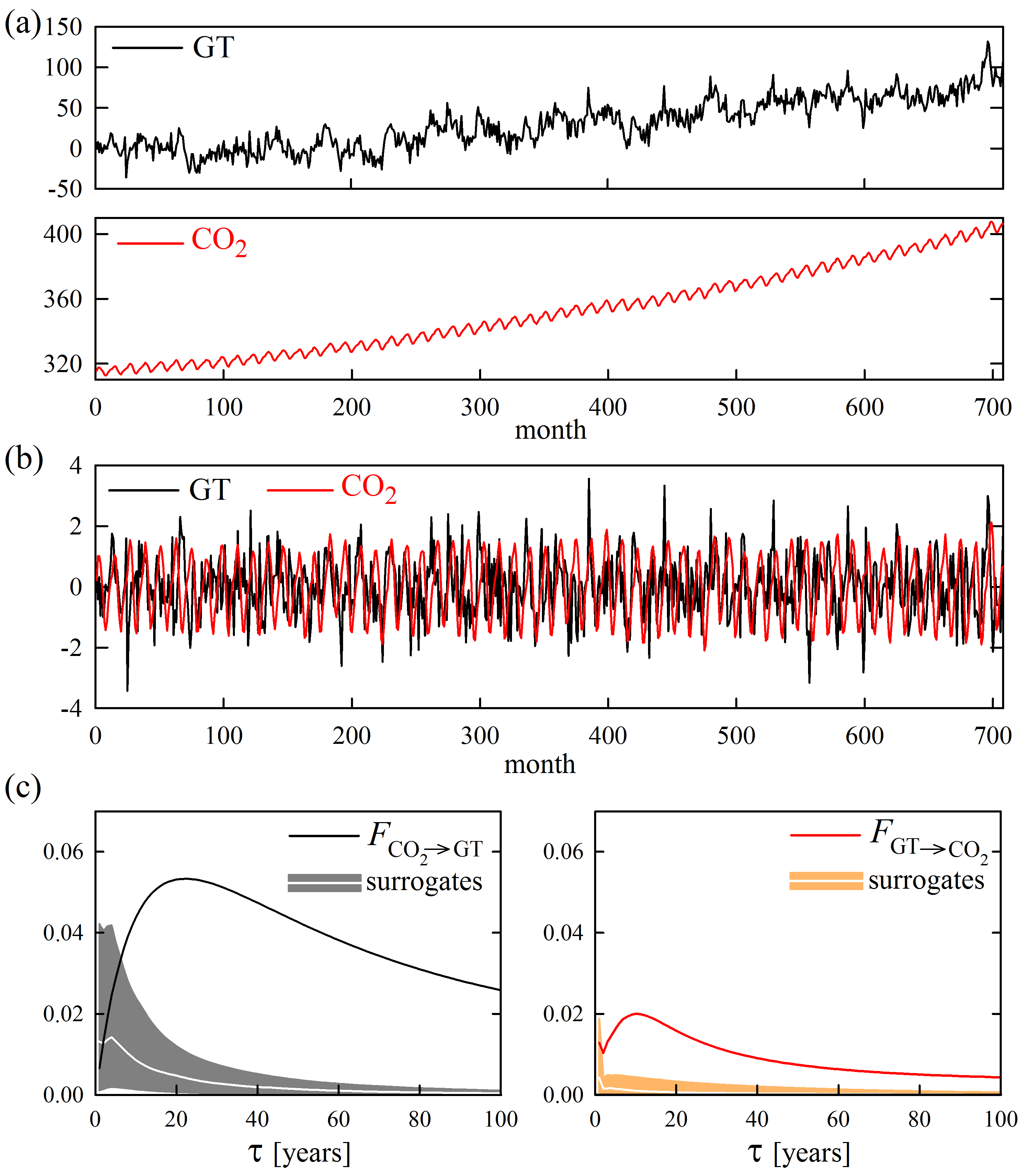}
\caption{\label{fig:climate1} Multiscale GC analysis of global temperature (GT) and  CO$_2$ concentration for modern climate data. Plots depict: (a) the original GT and CO$_2$  time series; (b) the time series superimposed after de-trending and normalization; and (c) the multiscale GC computed on the normalized time series (solid lines) and over 100 IAAFT surrogates (median: white lines; $5^{th}-95^{th}$ percentiles: shaded areas). Computations are performed using the proposed framework implemented with an order-6 FIR lowpass filter. The AR model order, set by the BIC criterion, is $p=14$.}
\end{figure}

Here, we first analyze the global land-ocean temperature index \cite{GISS} and $CO_2$ concentration \cite{CO2} measured at monthly resolution from  March 1958 to February 2017.
The measured time series, lasting 708 data points, are shown in Fig.~\ref{fig:climate1}(a). To fulfill stationarity criteria, we de-trended the two series applying an L1 norm filter. The analyzed time series, normalized to zero mean and unit variance, are shown in Fig.~\ref{fig:climate1}(b). We applied the proposed framework to compute GC along the two directions of interaction for time scales ranging from 1 to 100 years. To test the statistical significance of the estimated multiscale patterns of causality, the analysis was performed both for the original time series and for a set of 100 pairs of uncoupled time series sharing the autocorrelation and amplitude distribution of the original series; these surrogate series are generated using the iterative amplitude-adjusted Fourier transform (IAAFT) algorithm \cite{SchreiberIAAFT}.

The results depicted in Fig.~\ref{fig:climate1}(c) show that the GC along both directions is not distinguishable from the corresponding surrogate counterparts at $\tau$ equal one year, i.e. when standard GC analysis not encompassing multiple time scales is performed. On the other hand, the multiscale approach reveals, at longer time scales ($>$ 10 years), that GC is significantly higher than the surrogate threshold along both directions of interaction, thus showing the need of a multiscale approach to put in evidence this mutual interdependency between $CO_2$ and $GT$. Moreover we remark that the GC along the direction $CO_2 \to GT$ is characterized by a higher value of the statistics computed on the original time series, but also by higher values for the surrogate time series, compared with the direction $GT \to CO_2$.
\begin{figure}[ht!]
\includegraphics [width=8.5 cm] {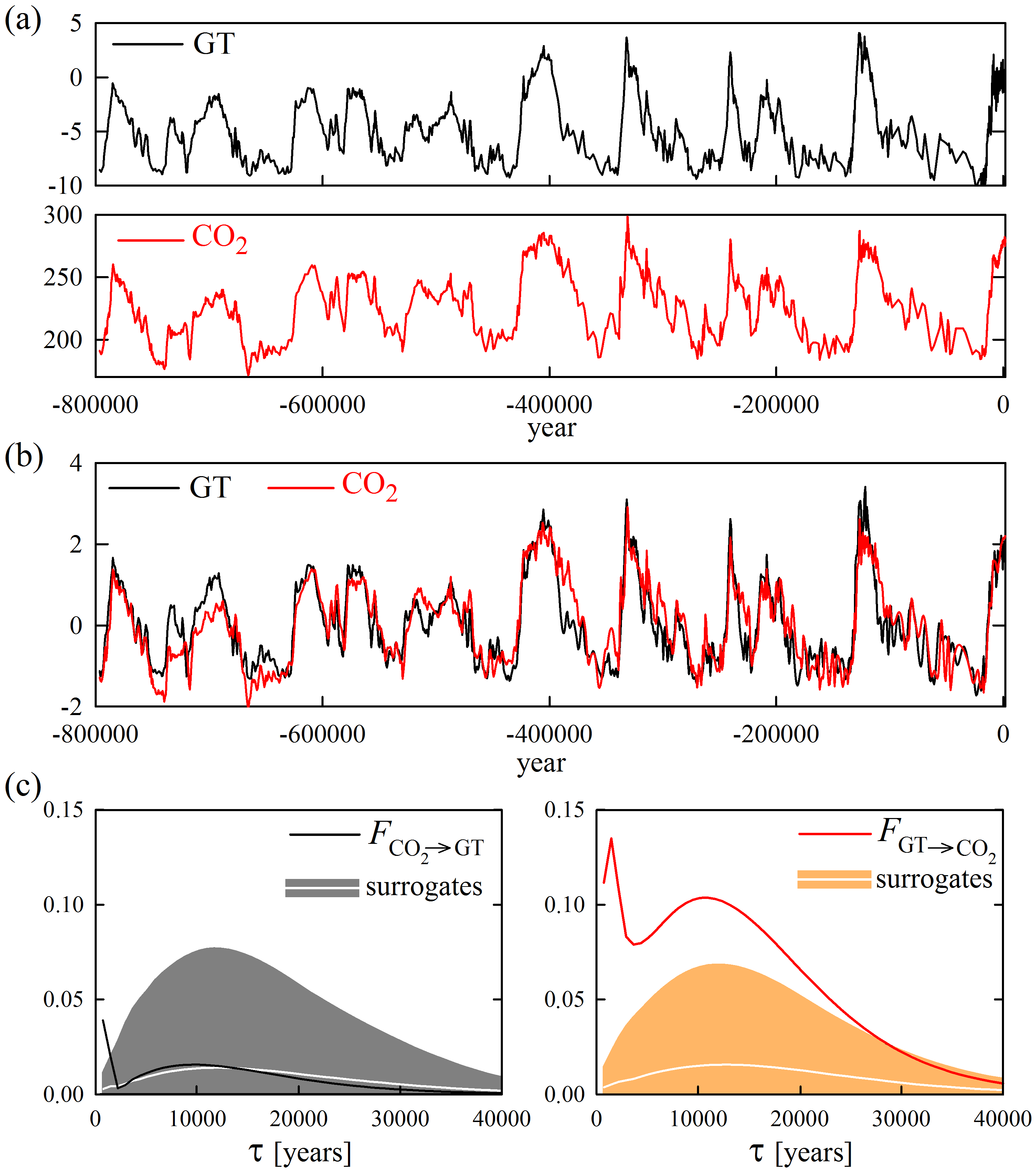}
\caption{\label{fig:climate2} Multiscale GC analysis of global temperature (GT) and  CO$_2$ concentration for paleoclimate data. Plots depict: (a) the original paleoclimatological time series; (b) the time series superimposed after uniform resampling of the time axis and normalization; and (c) the multiscale GC computed on the normalized time series (solid lines) and over 100 IAAFT surrogates (median: white lines; $5^{th}-95^{th}$ percentiles: shaded areas). Computations are performed using the proposed framework implemented with an order-6 FIR lowpass filter. The AR model order, set by the BIC criterion, is $p=3$.}
\end{figure}

Next, we dramatically change the time scales and turn to consider paleoclimatological data, so as to  analyze the GC  between  $GT$  and $CO_2$ concentration on the  Vostok  Ice  Core  data  from  400,000  to  6,000  years ago, extended by the EPICA Dome C data which go back to  800,000  years  ago \cite{ICECORE}.
The two time series, which are sampled with non-uniform time spacing, are shown in Fig.~\ref{fig:climate2}(a). Here, we studied the data resampled to an uniform time spacing of 729.77 years, corresponding to a time series length of 1095 points, and after normalization to zero mean and unit variance (Fig.~\ref{fig:climate2}(b)); results of multiscale GC analysis did not change substantially if the original non-uniformly sampled time series were considered, or applying slightly different uniform resampling.
In \cite{Kang2014}, empirical evidences for the existence of Granger causal influences along both directions 
$CO_2 \to GT$ and $GT \to CO_2$ have been found after correcting for deterministic trends on the same data.
Here, applying our framework for multiscale causality analysis we obtained the GC curves reported in Fig.~\ref{fig:climate2}(c). We find that, at paleolithic time scales, the GC $GT \to CO_2 $ is highly significant and peaks around 1000 and 10000 years. This result may be related to the lags between Antarctic deglacial warming and $CO_2$ increase reported in \cite{Caillon2003}, and also confirms the good evidences reported on the fact that higher global temperatures do promote a rise of greenhouse gas levels \cite{Scheffer2006}.
The opposite causal influence from $CO_2$ to GT is much less pronounced and exceeds the IAAFT threshold for statistical significance only at very small time scales.

Summarizing, our results show that carbon dioxide and temperature changes are interdependent at multiple time scales, with a predominance of $GT \to CO_2$ effects at paleolithic scales, and 
the presence of bidirectional causal interactions between $GT$ and $CO_2$ at the time scales of modern climate.
These results support the expectations that changing temperatures could be held responsible for changes in greenhouse gas concentrations on paleolithic time scales, while during the last 60 years the effect of human activities becomes evident as anthropogenic radiactive forcings are seemingly driving the global temperature changes. These causal relationships between $CO_2$ and global warming have been recently demonstrated in \cite{Stips2016}: in that work, the use of the rigorous formulation of information flow provided by \cite{XSLiang2014} led to evidence a clear unidirectional nature for the causal relation $GT \to CO_2$ in paleoclimatological data, and for the causal relation $CO_2 \to GT$ in modern climate data; in the same work, the application of the standard GC index to modern climate data suggested the presence of bidirectional effects $CO_2 \to GT$ and $GT \to CO_2$, thus pointing to some ambiguity in the assessment of a predominant direction of interaction using GC. Our results agree with this interpretation, as we do not find a prevalent causal direction using the classical GC index computed at the smallest time scale, and the use of surrogate data indicates the lack of statistical significance (Fig. 5). Nevertheless, the analysis performed at higher time scales
reveals the existence of significant GC $CO_2 \to GT$ and, for the first time to our knowledge, a nontrivial GC  $GT \to CO_2$ also in modern climate.
Although this result needs to be confirmed by the implementation of more robust measures of information flow, it may be of great relevance for climate studies as it is indicative of a positive feedback which will increase the effect of anthropogenic emissions on global temperatures.

\section{Conclusions} \label{sec:concl}
The present study makes the first step toward the theoretical understanding of multiscale causal relations between coupled  stochastic processes, and opens the way to the reliable estimation of these relations starting from simple AR identification. This will likely boost new impetus for research in the area of data-driven causality analysis, both in physics and in a wide variety of applicative fields.
The proposed framework is flexible enough to encompass more general model representations that may unveil important multiscale features of coupled processes. For instance, integrating the standard AR representation with fractional integrated (FI) innovation modeling \cite{sela2009computationally} would be straightforward as ARFI models have an SS representation, and would easily lead to assess multiscale GC in the presence of long-range correlations.

The proposed setting provides also the basis to expand the applicability of multiscale GC to nonstationary and nonlinear SS processes \cite{kitagawa1987non}, and to formalize exact computation of cross-scale information transfer within and between multivariate processes \cite{paluvs2014cross}, thus opening new avenues of research in the evaluation of causal interactions among coupled processes.
Of particular interest in this context is the recent formalization of the notion of information flow based on first principles, rather than axiomatic postulates or empirical proposals, implemented in \cite{XSLiang2016}. The latter work completes the rigorous formalism introduced in \cite{LIANG20071,LIANG2007173} and provides a well-principled alternative to the operational implementation of GC, and of transfer entropy intended as its non-parametric generalization, which are known to be complicated in many practical settings to an extent that spurious causalities may be revealed (e.g., in the presence of unobserved variables, measurement noise, or inappropriate time resolution) \cite{SmirnovSpurious, HahsPethel, NalatoreMitigating, ChicharroAlgorithms}.
Hence, the availability of a rigorous derivation of the information flowing among the components of discrete time stochastic mapping, provided in \cite{XSLiang2016} and extended therein to continuous time stochastic mappings and to deterministic systems, certainly constitutes a firm basis for the design of a more faithful analysis of causality between dynamical system components operated across multiple temporal scales.



\nocite{*}

\bibliography{MS_GC}

\begin{thebibliography}{48}%
\makeatletter
\providecommand \@ifxundefined [1]{%
 \@ifx{#1\undefined}
}%
\providecommand \@ifnum [1]{%
 \ifnum #1\expandafter \@firstoftwo
 \else \expandafter \@secondoftwo
 \fi
}%
\providecommand \@ifx [1]{%
 \ifx #1\expandafter \@firstoftwo
 \else \expandafter \@secondoftwo
 \fi
}%
\providecommand \natexlab [1]{#1}%
\providecommand \enquote  [1]{``#1''}%
\providecommand \bibnamefont  [1]{#1}%
\providecommand \bibfnamefont [1]{#1}%
\providecommand \citenamefont [1]{#1}%
\providecommand \href@noop [0]{\@secondoftwo}%
\providecommand \href [0]{\begingroup \@sanitize@url \@href}%
\providecommand \@href[1]{\@@startlink{#1}\@@href}%
\providecommand \@@href[1]{\endgroup#1\@@endlink}%
\providecommand \@sanitize@url [0]{\catcode `\\12\catcode `\$12\catcode
  `\&12\catcode `\#12\catcode `\^12\catcode `\_12\catcode `\%12\relax}%
\providecommand \@@startlink[1]{}%
\providecommand \@@endlink[0]{}%
\providecommand \url  [0]{\begingroup\@sanitize@url \@url }%
\providecommand \@url [1]{\endgroup\@href {#1}{\urlprefix }}%
\providecommand \urlprefix  [0]{URL }%
\providecommand \Eprint [0]{\href }%
\providecommand \doibase [0]{http://dx.doi.org/}%
\providecommand \selectlanguage [0]{\@gobble}%
\providecommand \bibinfo  [0]{\@secondoftwo}%
\providecommand \bibfield  [0]{\@secondoftwo}%
\providecommand \translation [1]{[#1]}%
\providecommand \BibitemOpen [0]{}%
\providecommand \bibitemStop [0]{}%
\providecommand \bibitemNoStop [0]{.\EOS\space}%
\providecommand \EOS [0]{\spacefactor3000\relax}%
\providecommand \BibitemShut  [1]{\csname bibitem#1\endcsname}%
\let\auto@bib@innerbib\@empty
\bibitem [{\citenamefont {Wiener}(1956)}]{wiener1956theory}%
  \BibitemOpen
  \bibfield  {author} {\bibinfo {author} {\bibfnamefont {N.}~\bibnamefont
  {Wiener}},\ }\href@noop {} {\bibfield  {journal} {\bibinfo  {journal} {Modern
  mathematics for engineers}\ }\textbf {\bibinfo {volume} {1}},\ \bibinfo
  {pages} {125} (\bibinfo {year} {1956})}\BibitemShut {NoStop}%
\bibitem [{\citenamefont {Granger}(1969)}]{granger1969}%
  \BibitemOpen
  \bibfield  {author} {\bibinfo {author} {\bibfnamefont {C.~W.}\ \bibnamefont
  {Granger}},\ }\href@noop {} {\bibfield  {journal} {\bibinfo  {journal}
  {Econometrica: Journal of the Econometric Society}\ ,\ \bibinfo {pages}
  {424}} (\bibinfo {year} {1969})}\BibitemShut {NoStop}%
\bibitem [{\citenamefont {Geweke}(1982)}]{geweke1982}%
  \BibitemOpen
  \bibfield  {author} {\bibinfo {author} {\bibfnamefont {J.}~\bibnamefont
  {Geweke}},\ }\href@noop {} {\bibfield  {journal} {\bibinfo  {journal}
  {Journal of the American statistical association}\ }\textbf {\bibinfo
  {volume} {77}},\ \bibinfo {pages} {304} (\bibinfo {year} {1982})}\BibitemShut
  {NoStop}%
\bibitem [{\citenamefont {Geweke}(1984)}]{geweke1984}%
  \BibitemOpen
  \bibfield  {author} {\bibinfo {author} {\bibfnamefont {J.~F.}\ \bibnamefont
  {Geweke}},\ }\href@noop {} {\bibfield  {journal} {\bibinfo  {journal} {J. Am.
  Stat. Ass.}\ }\textbf {\bibinfo {volume} {79}},\ \bibinfo {pages} {907}
  (\bibinfo {year} {1984})}\BibitemShut {NoStop}%
\bibitem [{\citenamefont {Freeman}(1983)}]{freeman1983granger}%
  \BibitemOpen
  \bibfield  {author} {\bibinfo {author} {\bibfnamefont {J.~R.}\ \bibnamefont
  {Freeman}},\ }\href@noop {} {\bibfield  {journal} {\bibinfo  {journal}
  {American Journal of Political Science}\ ,\ \bibinfo {pages} {327}} (\bibinfo
  {year} {1983})}\BibitemShut {NoStop}%
\bibitem [{\citenamefont {Smirnov}\ and\ \citenamefont
  {Mokhov}(2009)}]{Smirnov2009}%
  \BibitemOpen
  \bibfield  {author} {\bibinfo {author} {\bibfnamefont {D.~A.}\ \bibnamefont
  {Smirnov}}\ and\ \bibinfo {author} {\bibfnamefont {I.~I.}\ \bibnamefont
  {Mokhov}},\ }\href {\doibase 10.1103/PhysRevE.80.016208} {\bibfield
  {journal} {\bibinfo  {journal} {Physical Review E}\ }\textbf {\bibinfo
  {volume} {80}},\ \bibinfo {pages} {016208} (\bibinfo {year}
  {2009})}\BibitemShut {NoStop}%
\bibitem [{\citenamefont {Bressler}\ and\ \citenamefont
  {Seth}(2011)}]{bressler2011}%
  \BibitemOpen
  \bibfield  {author} {\bibinfo {author} {\bibfnamefont {S.~L.}\ \bibnamefont
  {Bressler}}\ and\ \bibinfo {author} {\bibfnamefont {A.~K.}\ \bibnamefont
  {Seth}},\ }\href@noop {} {\bibfield  {journal} {\bibinfo  {journal}
  {Neuroimage}\ }\textbf {\bibinfo {volume} {58}},\ \bibinfo {pages} {323}
  (\bibinfo {year} {2011})}\BibitemShut {NoStop}%
\bibitem [{\citenamefont {Porta}\ and\ \citenamefont {Faes}(2015)}]{Porta2015}%
  \BibitemOpen
  \bibfield  {author} {\bibinfo {author} {\bibfnamefont {A.}~\bibnamefont
  {Porta}}\ and\ \bibinfo {author} {\bibfnamefont {L.}~\bibnamefont {Faes}},\
  }\href {\doibase 10.1109/JPROC.2015.2476824} {\bibfield  {journal} {\bibinfo
  {journal} {Proceedings of the IEEE}\ } (\bibinfo {year} {2015}),\
  10.1109/JPROC.2015.2476824}\BibitemShut {NoStop}%
\bibitem [{\citenamefont {Barnett}\ \emph {et~al.}(2009)\citenamefont
  {Barnett}, \citenamefont {Barrett},\ and\ \citenamefont
  {Seth}}]{barnett2009granger}%
  \BibitemOpen
  \bibfield  {author} {\bibinfo {author} {\bibfnamefont {L.}~\bibnamefont
  {Barnett}}, \bibinfo {author} {\bibfnamefont {A.~B.}\ \bibnamefont
  {Barrett}}, \ and\ \bibinfo {author} {\bibfnamefont {A.~K.}\ \bibnamefont
  {Seth}},\ }\href@noop {} {\bibfield  {journal} {\bibinfo  {journal} {Phys.
  Rev. Lett.}\ }\textbf {\bibinfo {volume} {103}},\ \bibinfo {pages} {238701}
  (\bibinfo {year} {2009})}\BibitemShut {NoStop}%
\bibitem [{\citenamefont {Ivanov}\ \emph {et~al.}(1999)\citenamefont {Ivanov},
  \citenamefont {Nunes~Amaral}, \citenamefont {Goldberger}, \citenamefont
  {Havlin}, \citenamefont {Rosenblum}, \citenamefont {Struzik},\ and\
  \citenamefont {Stanley}}]{Ivanov1999461}%
  \BibitemOpen
  \bibfield  {author} {\bibinfo {author} {\bibfnamefont {P.}~\bibnamefont
  {Ivanov}}, \bibinfo {author} {\bibfnamefont {L.}~\bibnamefont
  {Nunes~Amaral}}, \bibinfo {author} {\bibfnamefont {A.}~\bibnamefont
  {Goldberger}}, \bibinfo {author} {\bibfnamefont {S.}~\bibnamefont {Havlin}},
  \bibinfo {author} {\bibfnamefont {M.}~\bibnamefont {Rosenblum}}, \bibinfo
  {author} {\bibfnamefont {Z.}~\bibnamefont {Struzik}}, \ and\ \bibinfo
  {author} {\bibfnamefont {H.}~\bibnamefont {Stanley}},\ }\href@noop {}
  {\bibfield  {journal} {\bibinfo  {journal} {Nature}\ }\textbf {\bibinfo
  {volume} {399}},\ \bibinfo {pages} {461} (\bibinfo {year}
  {1999})}\BibitemShut {NoStop}%
\bibitem [{\citenamefont {Kang}\ \emph {et~al.}(2009)\citenamefont {Kang},
  \citenamefont {Jia}, \citenamefont {Geocadin},\ and\ \citenamefont
  {Thakor}}]{Thakor2009}%
  \BibitemOpen
  \bibfield  {author} {\bibinfo {author} {\bibfnamefont {X.}~\bibnamefont
  {Kang}}, \bibinfo {author} {\bibfnamefont {X.}~\bibnamefont {Jia}}, \bibinfo
  {author} {\bibfnamefont {R.}~\bibnamefont {Geocadin}}, \ and\ \bibinfo
  {author} {\bibfnamefont {N.~V.}\ \bibnamefont {Thakor}},\ }\href@noop {}
  {\bibfield  {journal} {\bibinfo  {journal} {IEEE Trans. Biomed. Eng.}\
  }\textbf {\bibinfo {volume} {5}},\ \bibinfo {pages} {1023} (\bibinfo {year}
  {2009})}\BibitemShut {NoStop}%
\bibitem [{\citenamefont {Valencia}\ \emph {et~al.}(2009)\citenamefont
  {Valencia}, \citenamefont {Porta}, \citenamefont {Vallverd\'{u}},
  \citenamefont {Clari\'{a}}, \citenamefont {Baranowski}, \citenamefont
  {Or\l{}owska-Baranowska},\ and\ \citenamefont {Caminal}}]{Valencia20092202}%
  \BibitemOpen
  \bibfield  {author} {\bibinfo {author} {\bibfnamefont {J.}~\bibnamefont
  {Valencia}}, \bibinfo {author} {\bibfnamefont {A.}~\bibnamefont {Porta}},
  \bibinfo {author} {\bibfnamefont {M.}~\bibnamefont {Vallverd\'{u}}}, \bibinfo
  {author} {\bibfnamefont {F.}~\bibnamefont {Clari\'{a}}}, \bibinfo {author}
  {\bibfnamefont {R.}~\bibnamefont {Baranowski}}, \bibinfo {author}
  {\bibfnamefont {E.}~\bibnamefont {Or\l{}owska-Baranowska}}, \ and\ \bibinfo
  {author} {\bibfnamefont {P.}~\bibnamefont {Caminal}},\ }\href {\doibase
  10.1109/TBME.2009.2021986} {\bibfield  {journal} {\bibinfo  {journal} {IEEE
  Trans. Biomed. Eng.}\ }\textbf {\bibinfo {volume} {56}},\ \bibinfo {pages}
  {2202} (\bibinfo {year} {2009})}\BibitemShut {NoStop}%
\bibitem [{\citenamefont {Chou}(2011)}]{chou2011wavelet}%
  \BibitemOpen
  \bibfield  {author} {\bibinfo {author} {\bibfnamefont {C.-M.}\ \bibnamefont
  {Chou}},\ }\href@noop {} {\bibfield  {journal} {\bibinfo  {journal}
  {Entropy}\ }\textbf {\bibinfo {volume} {13}},\ \bibinfo {pages} {241}
  (\bibinfo {year} {2011})}\BibitemShut {NoStop}%
\bibitem [{\citenamefont {Wang}\ \emph {et~al.}(2013)\citenamefont {Wang},
  \citenamefont {Shang}, \citenamefont {Zhao},\ and\ \citenamefont
  {Xia}}]{wang2013multiscale}%
  \BibitemOpen
  \bibfield  {author} {\bibinfo {author} {\bibfnamefont {J.}~\bibnamefont
  {Wang}}, \bibinfo {author} {\bibfnamefont {P.}~\bibnamefont {Shang}},
  \bibinfo {author} {\bibfnamefont {X.}~\bibnamefont {Zhao}}, \ and\ \bibinfo
  {author} {\bibfnamefont {J.}~\bibnamefont {Xia}},\ }\href@noop {} {\bibfield
  {journal} {\bibinfo  {journal} {International Journal of Modern Physics C}\
  }\textbf {\bibinfo {volume} {24}},\ \bibinfo {pages} {1350006} (\bibinfo
  {year} {2013})}\BibitemShut {NoStop}%
\bibitem [{\citenamefont {Costa}\ \emph {et~al.}(2002)\citenamefont {Costa},
  \citenamefont {Goldberger},\ and\ \citenamefont
  {Peng}}]{costa2002multiscale}%
  \BibitemOpen
  \bibfield  {author} {\bibinfo {author} {\bibfnamefont {M.}~\bibnamefont
  {Costa}}, \bibinfo {author} {\bibfnamefont {A.~L.}\ \bibnamefont
  {Goldberger}}, \ and\ \bibinfo {author} {\bibfnamefont {C.-K.}\ \bibnamefont
  {Peng}},\ }\href@noop {} {\bibfield  {journal} {\bibinfo  {journal} {Phys.
  Rev. Lett.}\ }\textbf {\bibinfo {volume} {89}},\ \bibinfo {pages} {068102}
  (\bibinfo {year} {2002})}\BibitemShut {NoStop}%
\bibitem [{\citenamefont {Lungarella}\ \emph {et~al.}(2007)\citenamefont
  {Lungarella}, \citenamefont {Pitti},\ and\ \citenamefont
  {Kuniyoshi}}]{Lungarella2007}%
  \BibitemOpen
  \bibfield  {author} {\bibinfo {author} {\bibfnamefont {M.}~\bibnamefont
  {Lungarella}}, \bibinfo {author} {\bibfnamefont {A.}~\bibnamefont {Pitti}}, \
  and\ \bibinfo {author} {\bibfnamefont {Y.}~\bibnamefont {Kuniyoshi}},\
  }\href@noop {} {\bibfield  {journal} {\bibinfo  {journal} {Phys. Rev. E}\
  }\textbf {\bibinfo {volume} {76}} (\bibinfo {year} {2007})}\BibitemShut
  {NoStop}%
\bibitem [{\citenamefont {Palu{\v{s}}}(2014)}]{paluvs2014cross}%
  \BibitemOpen
  \bibfield  {author} {\bibinfo {author} {\bibfnamefont {M.}~\bibnamefont
  {Palu{\v{s}}}},\ }\href@noop {} {\bibfield  {journal} {\bibinfo  {journal}
  {Entropy}\ }\textbf {\bibinfo {volume} {16}},\ \bibinfo {pages} {5263}
  (\bibinfo {year} {2014})}\BibitemShut {NoStop}%
\bibitem [{\citenamefont {Barnett}\ and\ \citenamefont
  {Seth}(2015)}]{barnett2015granger}%
  \BibitemOpen
  \bibfield  {author} {\bibinfo {author} {\bibfnamefont {L.}~\bibnamefont
  {Barnett}}\ and\ \bibinfo {author} {\bibfnamefont {A.~K.}\ \bibnamefont
  {Seth}},\ }\href@noop {} {\bibfield  {journal} {\bibinfo  {journal} {Phys.
  Rev. E}\ }\textbf {\bibinfo {volume} {91}},\ \bibinfo {pages} {040101}
  (\bibinfo {year} {2015})}\BibitemShut {NoStop}%
\bibitem [{\citenamefont {Solo}(2016)}]{solo2016state}%
  \BibitemOpen
  \bibfield  {author} {\bibinfo {author} {\bibfnamefont {V.}~\bibnamefont
  {Solo}},\ }\href@noop {} {\bibfield  {journal} {\bibinfo  {journal} {Neural
  computation}\ }\textbf {\bibinfo {volume} {28}},\ \bibinfo {pages} {914}
  (\bibinfo {year} {2016})}\BibitemShut {NoStop}%
\bibitem [{\citenamefont {Barnett}\ and\ \citenamefont
  {Seth}(2011)}]{Barnett2011404}%
  \BibitemOpen
  \bibfield  {author} {\bibinfo {author} {\bibfnamefont {L.}~\bibnamefont
  {Barnett}}\ and\ \bibinfo {author} {\bibfnamefont {A.}~\bibnamefont {Seth}},\
  }\href {\doibase 10.1016/j.jneumeth.2011.08.010} {\bibfield  {journal}
  {\bibinfo  {journal} {J. Neurosci. Methods}\ }\textbf {\bibinfo {volume}
  {201}},\ \bibinfo {pages} {404} (\bibinfo {year} {2011})}\BibitemShut
  {NoStop}%
\bibitem [{\citenamefont {Aoki}\ and\ \citenamefont
  {Havenner}(1991)}]{Aoki1991}%
  \BibitemOpen
  \bibfield  {author} {\bibinfo {author} {\bibfnamefont {M.}~\bibnamefont
  {Aoki}}\ and\ \bibinfo {author} {\bibfnamefont {A.}~\bibnamefont
  {Havenner}},\ }\href {\doibase 10.1080/07474939108800194} {\bibfield
  {journal} {\bibinfo  {journal} {Econ. Rev.}\ }\textbf {\bibinfo {volume}
  {10}},\ \bibinfo {pages} {1} (\bibinfo {year} {1991})}\BibitemShut {NoStop}%
\bibitem [{\citenamefont {Anderson}\ and\ \citenamefont
  {Moore}(1979)}]{anderson1979optimal}%
  \BibitemOpen
  \bibfield  {author} {\bibinfo {author} {\bibfnamefont {B.~D.}\ \bibnamefont
  {Anderson}}\ and\ \bibinfo {author} {\bibfnamefont {J.~B.}\ \bibnamefont
  {Moore}},\ }\href@noop {} {\bibfield  {journal} {\bibinfo  {journal}
  {Englewood Cliffs}\ }\textbf {\bibinfo {volume} {21}},\ \bibinfo {pages} {22}
  (\bibinfo {year} {1979})}\BibitemShut {NoStop}%
\bibitem [{\citenamefont {Liang}(2016)}]{XSLiang2016}%
  \BibitemOpen
  \bibfield  {author} {\bibinfo {author} {\bibfnamefont {X.~S.}\ \bibnamefont
  {Liang}},\ }\href {\doibase 10.1103/PhysRevE.94.052201} {\bibfield  {journal}
  {\bibinfo  {journal} {Phys. Rev. E}\ }\textbf {\bibinfo {volume} {94}},\
  \bibinfo {pages} {052201} (\bibinfo {year} {2016})}\BibitemShut {NoStop}%
\bibitem [{\citenamefont {Oppenheim}\ and\ \citenamefont
  {Schafer}(1975)}]{oppenheimdigital}%
  \BibitemOpen
  \bibfield  {author} {\bibinfo {author} {\bibfnamefont {A.~V.}\ \bibnamefont
  {Oppenheim}}\ and\ \bibinfo {author} {\bibfnamefont {R.~W.}\ \bibnamefont
  {Schafer}},\ }\href@noop {} {\emph {\bibinfo {title} {Digital signal
  processing}}}\ (\bibinfo  {publisher} {Prentice-Hall Englewood Cliffs, NJ},\
  \bibinfo {year} {1975})\BibitemShut {NoStop}%
\bibitem [{\citenamefont {Liang}(2014)}]{XSLiang2014}%
  \BibitemOpen
  \bibfield  {author} {\bibinfo {author} {\bibfnamefont {X.~S.}\ \bibnamefont
  {Liang}},\ }\href {\doibase 10.1103/PhysRevE.90.052150} {\bibfield  {journal}
  {\bibinfo  {journal} {Phys. Rev. E}\ }\textbf {\bibinfo {volume} {90}},\
  \bibinfo {pages} {052150} (\bibinfo {year} {2014})}\BibitemShut {NoStop}%
\bibitem [{\citenamefont {Liang}(2015)}]{XSLiang2015}%
  \BibitemOpen
  \bibfield  {author} {\bibinfo {author} {\bibfnamefont {X.~S.}\ \bibnamefont
  {Liang}},\ }\href {\doibase 10.1103/PhysRevE.92.022126} {\bibfield  {journal}
  {\bibinfo  {journal} {Phys. Rev. E}\ }\textbf {\bibinfo {volume} {92}},\
  \bibinfo {pages} {022126} (\bibinfo {year} {2015})}\BibitemShut {NoStop}%
\bibitem [{\citenamefont {Faes}\ \emph {et~al.}(2016)\citenamefont {Faes},
  \citenamefont {Montalto}, \citenamefont {Stramaglia}, \citenamefont {Nollo},\
  and\ \citenamefont {Marinazzo}}]{faes2016multiscale}%
  \BibitemOpen
  \bibfield  {author} {\bibinfo {author} {\bibfnamefont {L.}~\bibnamefont
  {Faes}}, \bibinfo {author} {\bibfnamefont {A.}~\bibnamefont {Montalto}},
  \bibinfo {author} {\bibfnamefont {S.}~\bibnamefont {Stramaglia}}, \bibinfo
  {author} {\bibfnamefont {G.}~\bibnamefont {Nollo}}, \ and\ \bibinfo {author}
  {\bibfnamefont {D.}~\bibnamefont {Marinazzo}},\ }\href@noop {} {\bibfield
  {journal} {\bibinfo  {journal} {arXiv preprint arXiv:1602.06155}\ } (\bibinfo
  {year} {2016})}\BibitemShut {NoStop}%
\bibitem [{\citenamefont {Marple}(1987)}]{marple1987digital}%
  \BibitemOpen
  \bibfield  {author} {\bibinfo {author} {\bibfnamefont {S.~L.}\ \bibnamefont
  {Marple}},\ }\href@noop {} {\emph {\bibinfo {title} {Digital spectral
  analysis: with applications}}},\ Vol.~\bibinfo {volume} {5}\ (\bibinfo
  {publisher} {Prentice-Hall Englewood Cliffs, NJ},\ \bibinfo {year}
  {1987})\BibitemShut {NoStop}%
\bibitem [{\citenamefont {Booth}\ \emph {et~al.}(2012)\citenamefont {Booth},
  \citenamefont {Jones}, \citenamefont {Collins}, \citenamefont {Totterdell},
  \citenamefont {Cox}, \citenamefont {Sitch}, \citenamefont {Huntingford},
  \citenamefont {Betts}, \citenamefont {Harris},\ and\ \citenamefont
  {Lloyd}}]{Booth2012}%
  \BibitemOpen
  \bibfield  {author} {\bibinfo {author} {\bibfnamefont {B.~B.~B.}\
  \bibnamefont {Booth}}, \bibinfo {author} {\bibfnamefont {C.~D.}\ \bibnamefont
  {Jones}}, \bibinfo {author} {\bibfnamefont {M.}~\bibnamefont {Collins}},
  \bibinfo {author} {\bibfnamefont {I.~J.}\ \bibnamefont {Totterdell}},
  \bibinfo {author} {\bibfnamefont {P.~M.}\ \bibnamefont {Cox}}, \bibinfo
  {author} {\bibfnamefont {S.}~\bibnamefont {Sitch}}, \bibinfo {author}
  {\bibfnamefont {C.}~\bibnamefont {Huntingford}}, \bibinfo {author}
  {\bibfnamefont {R.~A.}\ \bibnamefont {Betts}}, \bibinfo {author}
  {\bibfnamefont {G.~R.}\ \bibnamefont {Harris}}, \ and\ \bibinfo {author}
  {\bibfnamefont {J.}~\bibnamefont {Lloyd}},\ }\href {\doibase
  10.1088/1748-9326/7/2/024002} {\bibfield  {journal} {\bibinfo  {journal}
  {Environmental Research Letters}\ }\textbf {\bibinfo {volume} {7}},\ \bibinfo
  {pages} {024002} (\bibinfo {year} {2012})}\BibitemShut {NoStop}%
\bibitem [{\citenamefont {Delworth}\ \emph {et~al.}(2006)\citenamefont
  {Delworth}, \citenamefont {Broccoli}, \citenamefont {Rosati}, \citenamefont
  {Stouffer}, \citenamefont {Balaji}, \citenamefont {Beesley}, \citenamefont
  {Cooke}, \citenamefont {Dixon}, \citenamefont {Dunne}, \citenamefont {Dunne}
  \emph {et~al.}}]{Delworth2006}%
  \BibitemOpen
  \bibfield  {author} {\bibinfo {author} {\bibfnamefont {T.~L.}\ \bibnamefont
  {Delworth}}, \bibinfo {author} {\bibfnamefont {A.~J.}\ \bibnamefont
  {Broccoli}}, \bibinfo {author} {\bibfnamefont {A.}~\bibnamefont {Rosati}},
  \bibinfo {author} {\bibfnamefont {R.~J.}\ \bibnamefont {Stouffer}}, \bibinfo
  {author} {\bibfnamefont {V.}~\bibnamefont {Balaji}}, \bibinfo {author}
  {\bibfnamefont {J.~A.}\ \bibnamefont {Beesley}}, \bibinfo {author}
  {\bibfnamefont {W.~F.}\ \bibnamefont {Cooke}}, \bibinfo {author}
  {\bibfnamefont {K.~W.}\ \bibnamefont {Dixon}}, \bibinfo {author}
  {\bibfnamefont {J.}~\bibnamefont {Dunne}}, \bibinfo {author} {\bibfnamefont
  {K.~A.}\ \bibnamefont {Dunne}},  \emph {et~al.},\ }\href {\doibase
  10.1175/JCLI3629.1} {\bibfield  {journal} {\bibinfo  {journal} {Journal of
  Climate}\ }\textbf {\bibinfo {volume} {19}},\ \bibinfo {pages} {643}
  (\bibinfo {year} {2006})}\BibitemShut {NoStop}%
\bibitem [{\citenamefont {Kodra}\ \emph {et~al.}(2011)\citenamefont {Kodra},
  \citenamefont {Chatterjee},\ and\ \citenamefont {Ganguly}}]{Kodra2011}%
  \BibitemOpen
  \bibfield  {author} {\bibinfo {author} {\bibfnamefont {E.}~\bibnamefont
  {Kodra}}, \bibinfo {author} {\bibfnamefont {S.}~\bibnamefont {Chatterjee}}, \
  and\ \bibinfo {author} {\bibfnamefont {A.~R.}\ \bibnamefont {Ganguly}},\
  }\href {\doibase 10.1007/s00704-010-0342-3} {\bibfield  {journal} {\bibinfo
  {journal} {Theoretical and Applied Climatology}\ }\textbf {\bibinfo {volume}
  {104}},\ \bibinfo {pages} {325} (\bibinfo {year} {2011})}\BibitemShut
  {NoStop}%
\bibitem [{\citenamefont {Attanasio}(2012)}]{Attanasio2012}%
  \BibitemOpen
  \bibfield  {author} {\bibinfo {author} {\bibfnamefont {A.}~\bibnamefont
  {Attanasio}},\ }\href {\doibase 10.1007/s00704-012-0634-x} {\bibfield
  {journal} {\bibinfo  {journal} {Theoretical and Applied Climatology}\
  }\textbf {\bibinfo {volume} {110}},\ \bibinfo {pages} {281} (\bibinfo {year}
  {2012})}\BibitemShut {NoStop}%
\bibitem [{\citenamefont {Stips}\ \emph {et~al.}(2016)\citenamefont {Stips},
  \citenamefont {Macias}, \citenamefont {Coughlan}, \citenamefont
  {Garcia-Gorriz},\ and\ \citenamefont {Liang}}]{Stips2016}%
  \BibitemOpen
  \bibfield  {author} {\bibinfo {author} {\bibfnamefont {A.}~\bibnamefont
  {Stips}}, \bibinfo {author} {\bibfnamefont {D.}~\bibnamefont {Macias}},
  \bibinfo {author} {\bibfnamefont {C.}~\bibnamefont {Coughlan}}, \bibinfo
  {author} {\bibfnamefont {E.}~\bibnamefont {Garcia-Gorriz}}, \ and\ \bibinfo
  {author} {\bibfnamefont {X.~S.}\ \bibnamefont {Liang}},\ }\href {\doibase
  10.1038/srep21691} {\bibfield  {journal} {\bibinfo  {journal} {Scientific
  Reports}\ }\textbf {\bibinfo {volume} {6}},\ \bibinfo {pages} {21691}
  (\bibinfo {year} {2016})}\BibitemShut {NoStop}%
\bibitem [{GIS()}]{GISS}%
  \BibitemOpen
  \href@noop {} {\enquote {\bibinfo {title} {Nasa - goddard institute for space
  studies},}\ }\bibinfo {howpublished} {\url{https://data.giss.nasa.gov/}},\
  \bibinfo {note} {accessed: 2017-03-23}\BibitemShut {NoStop}%
\bibitem [{CO2()}]{CO2}%
  \BibitemOpen
  \href@noop {} {\enquote {\bibinfo {title} {Earth system research
  laboratory},}\ }\bibinfo {howpublished} {\url{https://www.esrl.noaa.gov/}},\
  \bibinfo {note} {accessed: 2017-03-23}\BibitemShut {NoStop}%
\bibitem [{\citenamefont {Schreiber}\ and\ \citenamefont
  {Schmitz}(1996)}]{SchreiberIAAFT}%
  \BibitemOpen
  \bibfield  {author} {\bibinfo {author} {\bibfnamefont {T.}~\bibnamefont
  {Schreiber}}\ and\ \bibinfo {author} {\bibfnamefont {A.}~\bibnamefont
  {Schmitz}},\ }\href {\doibase 10.1103/PhysRevLett.77.635} {\bibfield
  {journal} {\bibinfo  {journal} {Phys. Rev. Lett.}\ }\textbf {\bibinfo
  {volume} {77}},\ \bibinfo {pages} {635} (\bibinfo {year} {1996})}\BibitemShut
  {NoStop}%
\bibitem [{ICE()}]{ICECORE}%
  \BibitemOpen
  \href@noop {} {\enquote {\bibinfo {title} {National oceanic and atmospheric
  administration},}\ }\bibinfo {howpublished}
  {\url{https://www.ncdc.noaa.gov/data-access/paleoclimatology-data/datasets/ice-core}},\
  \bibinfo {note} {accessed: 2017-03-23}\BibitemShut {NoStop}%
\bibitem [{\citenamefont {Kang}\ and\ \citenamefont
  {Larsson}(2014)}]{Kang2014}%
  \BibitemOpen
  \bibfield  {author} {\bibinfo {author} {\bibfnamefont {J.}~\bibnamefont
  {Kang}}\ and\ \bibinfo {author} {\bibfnamefont {R.}~\bibnamefont {Larsson}},\
  }\href {\doibase 10.1007/s00704-013-0960-7} {\bibfield  {journal} {\bibinfo
  {journal} {Theoretical and Applied Climatology}\ }\textbf {\bibinfo {volume}
  {116}},\ \bibinfo {pages} {537} (\bibinfo {year} {2014})}\BibitemShut
  {NoStop}%
\bibitem [{\citenamefont {Caillon}(2003)}]{Caillon2003}%
  \BibitemOpen
  \bibfield  {author} {\bibinfo {author} {\bibfnamefont {N.}~\bibnamefont
  {Caillon}},\ }\href {\doibase 10.1126/science.1078758} {\bibfield  {journal}
  {\bibinfo  {journal} {Science}\ }\textbf {\bibinfo {volume} {299}},\ \bibinfo
  {pages} {1728} (\bibinfo {year} {2003})}\BibitemShut {NoStop}%
\bibitem [{\citenamefont {Scheffer}\ \emph {et~al.}(2006)\citenamefont
  {Scheffer}, \citenamefont {Brovkin},\ and\ \citenamefont
  {Cox}}]{Scheffer2006}%
  \BibitemOpen
  \bibfield  {author} {\bibinfo {author} {\bibfnamefont {M.}~\bibnamefont
  {Scheffer}}, \bibinfo {author} {\bibfnamefont {V.}~\bibnamefont {Brovkin}}, \
  and\ \bibinfo {author} {\bibfnamefont {P.~M.}\ \bibnamefont {Cox}},\ }\href
  {\doibase 10.1029/2005GL025044} {\bibfield  {journal} {\bibinfo  {journal}
  {Geophysical Research Letters}\ }\textbf {\bibinfo {volume} {33}},\ \bibinfo
  {pages} {L10702} (\bibinfo {year} {2006})}\BibitemShut {NoStop}%
\bibitem [{\citenamefont {Sela}\ and\ \citenamefont
  {Hurvich}(2009)}]{sela2009computationally}%
  \BibitemOpen
  \bibfield  {author} {\bibinfo {author} {\bibfnamefont {R.~J.}\ \bibnamefont
  {Sela}}\ and\ \bibinfo {author} {\bibfnamefont {C.~M.}\ \bibnamefont
  {Hurvich}},\ }\href@noop {} {\bibfield  {journal} {\bibinfo  {journal}
  {Journal of Time Series Analysis}\ }\textbf {\bibinfo {volume} {30}},\
  \bibinfo {pages} {631} (\bibinfo {year} {2009})}\BibitemShut {NoStop}%
\bibitem [{\citenamefont {Kitagawa}(1987)}]{kitagawa1987non}%
  \BibitemOpen
  \bibfield  {author} {\bibinfo {author} {\bibfnamefont {G.}~\bibnamefont
  {Kitagawa}},\ }\href@noop {} {\bibfield  {journal} {\bibinfo  {journal}
  {Journal of the American statistical association}\ }\textbf {\bibinfo
  {volume} {82}},\ \bibinfo {pages} {1032} (\bibinfo {year}
  {1987})}\BibitemShut {NoStop}%
\bibitem [{\citenamefont {Liang}\ and\ \citenamefont
  {Kleeman}(2007{\natexlab{a}})}]{LIANG20071}%
  \BibitemOpen
  \bibfield  {author} {\bibinfo {author} {\bibfnamefont {X.~S.}\ \bibnamefont
  {Liang}}\ and\ \bibinfo {author} {\bibfnamefont {R.}~\bibnamefont
  {Kleeman}},\ }\href {\doibase http://dx.doi.org/10.1016/j.physd.2007.04.002}
  {\bibfield  {journal} {\bibinfo  {journal} {Physica D: Nonlinear Phenomena}\
  }\textbf {\bibinfo {volume} {231}},\ \bibinfo {pages} {1 } (\bibinfo {year}
  {2007}{\natexlab{a}})}\BibitemShut {NoStop}%
\bibitem [{\citenamefont {Liang}\ and\ \citenamefont
  {Kleeman}(2007{\natexlab{b}})}]{LIANG2007173}%
  \BibitemOpen
  \bibfield  {author} {\bibinfo {author} {\bibfnamefont {X.~S.}\ \bibnamefont
  {Liang}}\ and\ \bibinfo {author} {\bibfnamefont {R.}~\bibnamefont
  {Kleeman}},\ }\href {\doibase http://dx.doi.org/10.1016/j.physd.2006.12.012}
  {\bibfield  {journal} {\bibinfo  {journal} {Physica D: Nonlinear Phenomena}\
  }\textbf {\bibinfo {volume} {227}},\ \bibinfo {pages} {173 } (\bibinfo {year}
  {2007}{\natexlab{b}})}\BibitemShut {NoStop}%
\bibitem [{\citenamefont {Smirnov}(2013)}]{SmirnovSpurious}%
  \BibitemOpen
  \bibfield  {author} {\bibinfo {author} {\bibfnamefont {D.~A.}\ \bibnamefont
  {Smirnov}},\ }\href {\doibase 10.1103/PhysRevE.87.042917} {\bibfield
  {journal} {\bibinfo  {journal} {Phys. Rev. E}\ }\textbf {\bibinfo {volume}
  {87}},\ \bibinfo {pages} {042917} (\bibinfo {year} {2013})}\BibitemShut
  {NoStop}%
\bibitem [{\citenamefont {Hahs}\ and\ \citenamefont
  {Pethel}(2011)}]{HahsPethel}%
  \BibitemOpen
  \bibfield  {author} {\bibinfo {author} {\bibfnamefont {D.~W.}\ \bibnamefont
  {Hahs}}\ and\ \bibinfo {author} {\bibfnamefont {S.~D.}\ \bibnamefont
  {Pethel}},\ }\href {\doibase 10.1103/PhysRevLett.107.128701} {\bibfield
  {journal} {\bibinfo  {journal} {Phys. Rev. Lett.}\ }\textbf {\bibinfo
  {volume} {107}},\ \bibinfo {pages} {128701} (\bibinfo {year}
  {2011})}\BibitemShut {NoStop}%
\bibitem [{\citenamefont {Nalatore}\ \emph {et~al.}(2007)\citenamefont
  {Nalatore}, \citenamefont {Ding},\ and\ \citenamefont
  {Rangarajan}}]{NalatoreMitigating}%
  \BibitemOpen
  \bibfield  {author} {\bibinfo {author} {\bibfnamefont {H.}~\bibnamefont
  {Nalatore}}, \bibinfo {author} {\bibfnamefont {M.}~\bibnamefont {Ding}}, \
  and\ \bibinfo {author} {\bibfnamefont {G.}~\bibnamefont {Rangarajan}},\
  }\href {\doibase 10.1103/PhysRevE.75.031123} {\bibfield  {journal} {\bibinfo
  {journal} {Phys. Rev. E}\ }\textbf {\bibinfo {volume} {75}},\ \bibinfo
  {pages} {031123} (\bibinfo {year} {2007})}\BibitemShut {NoStop}%
\bibitem [{\citenamefont {Chicharro}\ and\ \citenamefont
  {Panzeri}(2014)}]{ChicharroAlgorithms}%
  \BibitemOpen
  \bibfield  {author} {\bibinfo {author} {\bibfnamefont {D.}~\bibnamefont
  {Chicharro}}\ and\ \bibinfo {author} {\bibfnamefont {S.}~\bibnamefont
  {Panzeri}},\ }\href {\doibase 10.3389/fninf.2014.00064} {\bibfield  {journal}
  {\bibinfo  {journal} {Frontiers in Neuroinformatics}\ }\textbf {\bibinfo
  {volume} {8}},\ \bibinfo {pages} {64} (\bibinfo {year} {2014})}\BibitemShut
  {NoStop}%
\end{thebibliography}%

\end{document}